\documentclass[aps,prl,twocolumn,showpacs,preprintnumbers,
amsmath,amssymb,floatfix,nofootinbib,reprint]{revtex4}

\usepackage{hyperref}
\usepackage{amsmath}
\usepackage{amsfonts}
\usepackage{amssymb}
\usepackage{graphicx}
\usepackage{color}
\usepackage{bm}

\usepackage[caption=false, font=small]{subfig}
\usepackage[utf8]{inputenc}

\begin{document}

%
%
%
%

%
%

\title{Logical operations with single x-ray photons via dynamically-controlled nuclear resonances}

%
%

\author{Jonas \surname{Gunst}}
\email{Jonas.Gunst@mpi-hd.mpg.de}
\affiliation{Max-Planck-Institut f\"ur Kernphysik, Saupfercheckweg 1, D-69117 Heidelberg, Germany}

\author{Christoph H. \surname{Keitel}}
\affiliation{Max-Planck-Institut f\"ur Kernphysik, Saupfercheckweg 1, D-69117 Heidelberg, Germany}

\author{Adriana \surname{P\'alffy}}
\email{Palffy@mpi-hd.mpg.de}
\affiliation{Max-Planck-Institut f\"ur Kernphysik, Saupfercheckweg 1, D-69117 Heidelberg, Germany}


\date{\today}

%
%
%
%
%
%
%
\begin{abstract}

The implementation of logical operations on polarization-encoded x-rays via resonant light-nucleus interactions is theoretically investigated. We show that by means of resonant scattering off nuclei and fast rotations of the nuclear hyperfine magnetic field to control 
the polarization of the output photon, single-qubit logical gates can be simulated.  A second control qubit may be employed to trigger the magnetic field rotation, thus allowing several implementation choices for a controlled NOT gate for x-ray photons. %

\end{abstract}


\maketitle

While atomic transitions are naturally used to resonantly manipulate optical photons, nuclear transitions may be the elementary counterparts for x-rays. A number of fundamental phenomena such as for instance electromagnetically induced transparency \cite{Rohlsberger2012.EIT},  collective Lamb shift \cite{Rohlsberger2010.coll-Lamb}, slow light \cite{Olga2009.slow-light, Kilian.slow-light}, spontaneously generated coherence \cite{Kilian.atomic-coherences} or single-photon revival \cite{Olga2013.gamma-revival} has already been experimentally transferred to the nuclear realm. Coherent self-seeded x-ray free-electron lasers in development today \cite{Amann2012.self-seeded-XFEL} are anticipated to further promote nuclear quantum optics \cite{Adams2013.JoMO, Burvenich2006.PRL, Palffy2008.PRC} aiming at long-term objectives like $\gamma$-ray lasing \cite{Baldwin.gamma-laser, Olga1999.opt-control-mossbauer,Coussement2002.PRL} or controlled energy storage in nuclear metastable states \cite{Walker1999.N,Palffy2007.PRL,Jonas.Mo-triggering}. For quantum optics applications, nuclear transitions present a clean, well isolated system with very long coherence times, while x-rays attract with their good detection efficiency, penetration power and remarkable focus. Admittedly, experimental challenges at the large coherent x-ray source facilities today will require a different paradigm compared to table-top optical experiments. However,  control at the single-photon level has been recently demonstrated also in a laboratory-scale M\"ossbauer setup, where  the coherent manipulation of  waveforms of individual x-rays has been achieved \cite{Olga2014.N}. Such control procedures operated at single-photon nuclear interfaces open the perspective to extend fields like quantum information and quantum communication to photon energies in the keV-range.

The elementary building block of all quantum information protocols is the information carrier, the so-called qubit \cite{NielsenChuang2000}. X-ray photonic qubits potentially have sub-\r{A} spatial resolution \cite{Doring2013.OE},  drastically reducing the fundamental limitation on nanoscale photonic circuits. A promising way to encode information in single x-rays is to employ orthogonal polarization states like it is accomplished in the optical regime \cite{Kok2007.RMP, Kocsis2013.NP, Crespi2011.NC, OBrien2003.N}. X-ray linear polarization can be  measured with  precision up to $0.3^{\circ}$ using polarimeters based on the Compton effect \cite{Tashenov2011.PRL, Tashenov2013.PRA}, and Bragg reflections on crystals can filter polarizations states as good as $10^{-6}\%$ \cite{Toellner1995.polarizer,Marx2011.polarizer}. However, such information encoding requires precise control and processing schemes for the polarization of individual x-rays so far not addressed.

In this work, we investigate theoretically how polarization-encoded single x-rays can be coherently processed by means of resonant nuclear interactions. A broadband x-ray pulse resonant to a nuclear transition impinges on a target in the presence of a hyperfine magnetic field and produces a single nuclear excitation. Fast rotations of the hyperfine field \cite{Shvydko1996.PRL} are used to actively manipulate the polarization of the single-photon response of the nuclear target.  We show that it is feasible to implement one-qubit logical gates via such magnetic field rotations, and even binary gates by introducing in addition a second, temporally synchronized control photon. In particular, a possible x-ray photonic realization of the destructive controlled NOT (C-NOT) gate is put forward.

\begin{figure}%
\includegraphics[width=1.0\linewidth]{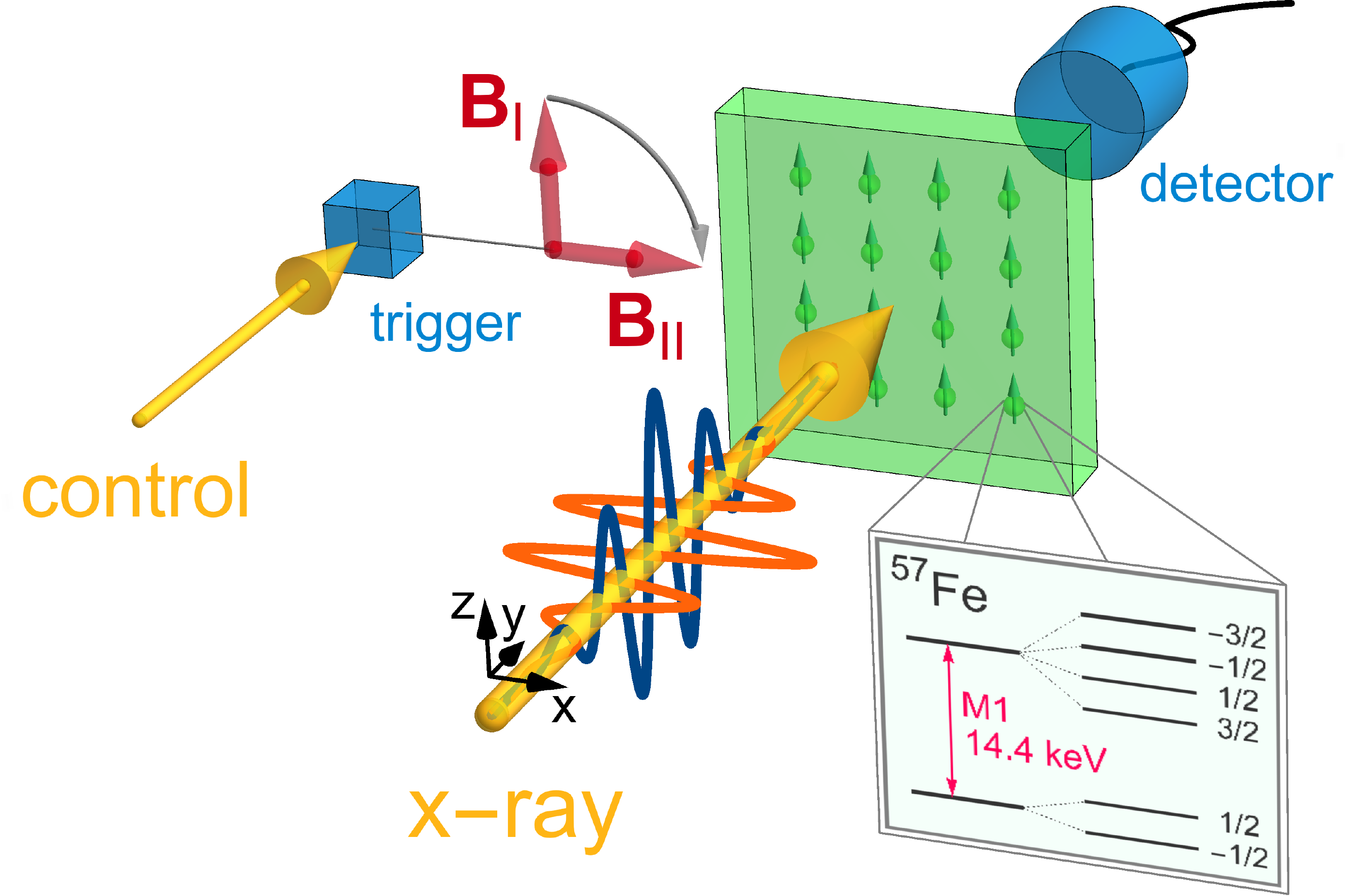}%
\caption{(color online) Nuclear forward scattering setup. $\sigma$- (orange, lighter hue) or $\pi$-polarized (blue, darker hue) x-rays scatter off a nuclear target in the forward direction. A
spatially separated control photon triggers a magnetic field rotation from the $z$- to the $x$-axis. The hyperfine-split nuclear level scheme of $^{57}$Fe is illustrated in the inset.}%
\label{fig:setup}%
\end{figure}

At present, typically in x-ray-nuclear-transition interfaces  only one excitation, i.e., one resonant x-ray photon, exists in the system at any given time \cite{Roehlsberger2004}. The photon is  in a coherent state rather than a Fock state, allowing for a semi-classical treatment of the light-nucleus interaction \cite{ScullyZubairy}.  Entanglement  occurs in the system only in its single-photon version \cite{Lee2000,palffy2009}, by having the x-ray photon entangle two spatially or temporally separated field modes. Furthermore, in contrast to nuclear magnetic resonance techniques \cite{NielsenChuang2000} that employ nuclear ground state spins as information carriers and process them with microwave fields, here we envisage magnetic fields to modify the properties of an x-ray transition to an excited nuclear state. These are rather unusual factors for the implementation of logical operations as known from the optical  \cite{Jaksch2000.PRL,Calarco2000.PRA,Calarco2003.PRA,Hammerer2010.review,Halfmann2011.PRA} or microwave regimes \cite{Cory1997.NMR, Gershenfeld1997.NMR,Wunderlich2001.PRL,Wunderlich2011.Nature}, that call for a new approach.

The x-ray-nucleus interaction is considered here in a nuclear forward scattering (NFS) setup  as  presented in  Fig.~\ref{fig:setup}.  The x-rays, typically from a synchrotron radiation (SR) source monochromatized to the nuclear transition energy, propagate along the $y$-direction and impinge on the nuclear sample with an incident angle of $90^{\circ}$. The radiation is linearly polarized with $x$-($z$-)polarized light denoted as $\sigma$-($\pi$-)polarization by convention \cite{Siddons1999.HI}. In quantum information $\sigma$ is often referred to as horizontal polarization (H) and $\pi$ as vertical (V). The time spectrum of the resonantly scattered radiation is detected in the forward direction. The nuclear response occurs on a much longer time scale than the x-ray pulse duration and the non-resonant, electronic response, allowing for time gating of the signal \cite{Roehlsberger2004}. Due to the typically narrow nuclear resonances and the low brilliance of x-ray sources, at most one nucleus can be excited in the sample. The most used nuclear transition with the NFS technique is the one connecting the stable ground state of $^{57}$Fe (nuclear spin $I_g=1/2$) with the the first excited state (nuclear spin $I_e=3/2$, mean lifetime $\tau$=141~ns) at 14.413 keV.  The recoilless nature of this transition in solid-state nuclear targets leads to the formation of a delocalized, collective excitation (in literature referred to as  nuclear exciton \cite{Hannon1999.HI} or  timed Dicke state \cite{Scully2009.superradiance}) which decays coherently  into the forward direction  leading to a relative speed-up and enhancement of the NFS yield \cite{Hastings1991.PRL, Smirnov1996.HI, vanBurck1987.PRL}.

In the presence of a nuclear hyperfine magnetic field, the ground and excited nuclear states undergo Zeeman splitting according to their spin values as illustrated in the inset of Fig.~\ref{fig:setup}. Due to the Fourier limit of the temporally narrow incident x-ray pulses, several polarization-selected hyperfine transitions labeled in the following with the index $l$ can be simultaneously driven  leading to well-known quantum beats in the NFS intensity spectrum \cite{Hannon1999.HI}. For instance, initially $\sigma$-($\pi$-)polarized x-rays couple to all $\Delta$m=0 ($\Delta$m=$\pm$1) transitions provided the magnetic field $\bm{B}$ points along the $z$-direction. Since only those photons are coherently scattered into the forward direction for which the nucleus returns  to its original ground state Zeeman level, the $\sigma$-($\pi$-)polarization is conserved in the course of NFS with constant hyperfine field $\bm{B}_{\mathrm{I}}$.

Abrupt rotations of the nuclear hyperfine magnetic field offer means of polarization control of the nuclear signal.  The magnetic field at the nuclear target  $\bm{B}_{\mathrm{I}}$ is initially assumed to be constant and to point along the $z$-axis. A fast rotation of the magnetic field after the nuclear excitation has taken place (for instance, to $\bm{B}_{\mathrm{II}}$ parallel to the $x$-axis by a 90${}^{\circ}$ counterclockwise rotation around the $y$ direction) leads to an almost instantaneous change of the quantization axis and a redistribution of the collective excitation among the Zeeman levels. Each initially excited nuclear current \cite{Shvydko1999.PRB,palffyjomo} is transferred into a sextet of new currents which can interfere constructively or destructively depending on the switching geometry and exact rotation moment $t_0$ \cite{Shvydko1996.PRL}. Experimentally such abrupt rotations could be realized on a timescale faster than 4~ns \cite{Shvydko1994.EPL} in $^{57}$Fe-enriched FeBO${}_3$ \cite{vanBurck1987.PRL,Smirnov1983.JETP} due to the special magnetization properties of the latter.

We describe the coherent nuclear scattering process by a semi-classical wave equation following the procedure presented in \cite{Shvydko1999.PRB}. 
The x-ray electric field in front and behind the target can be written as a time-modulated plane wave $\bm{E}(y,t) \text{e}^{\text{i} (k y - \omega t)}$. 
The calculation of the scattered field amplitude behind the target is carried out within the slowly-varying envelope approximation using perturbation theory and can be written as a summation over all multiple scattering orders $p$ from 1 to $\infty$. The  incident pulse $p$=0 is not of interest here and is typically eliminated in experiments by means of time gating. In a first approximation all multiple scattering events are assumed to occur only before the magnetic field switching, leading to the following expression for the electric field 
\begin{equation}
\bm{E}(\xi,t) \propto \sum_{p=1}^{\infty} \sum_l \frac{(-\xi)^{p}}{p!} \bm{\mathcal{A}}^{(p)}_l(t_0,\alpha,\beta,\gamma)\, \text{e}^{-\text{i} \Delta_l t - \Gamma_0 t/2}.
\label{eq:field}
\end{equation}
Here, $\xi$ is the optical depth of the medium, $\Delta_l$ describes the detuning from the nuclear transition frequency $\omega_0$ due to magnetic hyperfine splitting and $\Gamma_0$ represents the natural transition width. 
For each contributing  nuclear transition $l$ between the hyperfine-split levels and scattering order $p$, the time-independent amplitudes $\bm{\mathcal{A}}^{(p)}_l$ are completely determined by the magnetic field rotation geometry via the Euler angles $\alpha$, $\beta$ and $\gamma$ and by the switching time $t_0$. The expression (\ref{eq:field}) represents the dominating contribution to the scattered field and can be used to determine up to a good approximation the desired switching parameters.
By changing the order of the summations, the scattered radiation via the nuclear transition $l$ can be expressed as a product between a sum of time-independent amplitudes and a time-dependent phase factor, with specific parameter sets  for which constructive or destructive interference between the summation terms with different $l$ occur. A suitable choice of $t_0$, $\alpha$, $\beta$ and $\gamma$ can  control the scattered photon polarization on single-photon nuclear interfaces,  building the basis for the compilation of logical x-ray gates.

Classically, there are four one-qubit, i.e, unary, logic gates: the \textit{identity} leaves the target bit unchanged; the \textit{true} and \textit{false} operations give ``1'' and ``0'', respectively, independent of the input; and the \textit{negation} flips the operated bit from either ``0'' to ``1'' or from ``1'' to ``0''.  The single-photon qubits can be encoded as x-ray orthogonal polarization states, for instance  ``0'' as $\pi$- and ``1'' as $\sigma$-polarization.  The unary gates for polarization-encoded x-rays can be implemented  in resonant NFS by a timed $90^{\circ}$-rotation (from $z$ to $x$) of the magnetic field as shown in Fig.~\ref{fig:setup}.  Practically this corresponds to finding switching instances $t_0$ where $\sigma$- and $\pi$-polarizations are simultaneously converted  into (pure) opposite polarization states.

\begin{figure}
\includegraphics[width=1.0\linewidth]{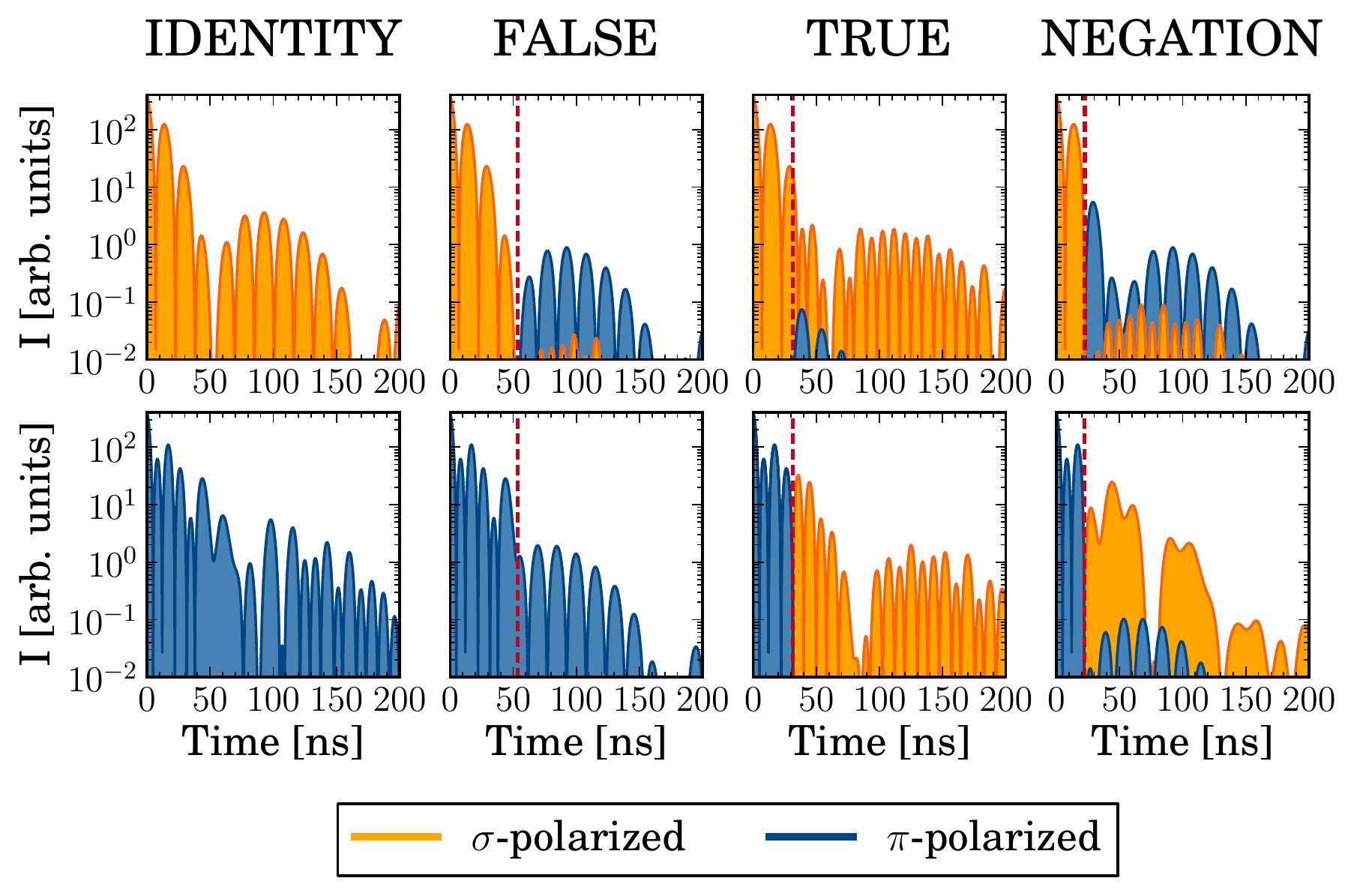}%
\caption{(color online). NFS intensity spectra with an optical depth of $\xi$=10 are shown for initially $\sigma$- (top row) and $\pi-$polarized (bottom row) x-rays. The switching times $t_0$ (red dashed lines) determine the implemented logical operation.}%
\label{fig:unary-gates}%
\end{figure}

Employing Eq.~(\ref{eq:field}) we can prove that such almost simultaneous switching times exist, allowing the implementation of the unary x-ray gates within the same setup. The numerical results for the scattered field are obtained going beyond the approximation in Eq.~(\ref{eq:field}) to include all multiple scattering events before and after $t_0$. The sum over the scattering order $p$ converges quickly such that including the first 14 scattering orders ($p_{max}$=14)  is already sufficient.
 Our results are presented in Fig.~\ref{fig:unary-gates}. In the top row the incident radiation is $\sigma$-polarized (orange line and filling, lighter hue) whereas the bottom row shows the scattered photon yield for initially $\pi$-polarized x-rays (blue line and filling, darker hue). Fig.~\ref{fig:unary-gates} shows that depending on the moments of magnetic field rotation, it is possible to convert orthogonal polarizations into each other in the course of NFS. For instance, a magnetic field rotation of $90^{\circ}$ at t$\approx$22.3~ns simultaneously converts $\sigma$ into $\pi$ and vice versa as shown in the last column of Fig.~\ref{fig:unary-gates}, successfully implementing the logical negation for all times $t>t_0$. In the same manner, it is also possible to realize the true, false and identity operation. The latter does not need any switching of the magnetic field  since the chosen geometry conserves $\sigma$- and $\pi$-polarization in the case of static hyperfine fields.

The polarization purity of the scattered radiation after the magnetic field switching is limited by three factors.  First, in our choice of $t_0$ we rely on Eq.~(\ref{eq:field}) in which we have disregarded the possibility of further multiple scattering after the magnetic field switching at $t_0$. This approximation leads to small polarization mixing as shown by the complete numerical calculation, see Fig.~\ref{fig:unary-gates}. Thereby, the unwanted polarization component contributes less than 4\% to the total intensity after $t_0$. Second, theoretically the switching times for an individual unary gate may differ by up to 0.5 ns depending on whether the incident radiation is $\sigma$- or $\pi$ polarized. For instance, in the case of the negation unary gate, $t_0^{\sigma}$=22.6~ns and  $t_0^{\pi}$=22.1~ns (these are the values used in Fig.~\ref{fig:unary-gates}). Even if we choose an averaged switching time in between, the calculated probability of realization for times $t>t_0$ is still better than 95\% for all four unary gates. Third, so far experimentally the switching time is known only to be less than 4 ns \cite{Shvydko1994.EPL} for a setup employing $^{57}$Fe-enriched FeBO${}_3$ \cite{vanBurck1987.PRL,Smirnov1983.JETP}. However, the good performance of switching experiments \cite{Shvydko1994.HI,Shvydko1996.PRL} indicates that the magnetic field rotation occurs on a shorter time than 4~ns. Already a rotation duration of 1 ns, leading to 1 ns uncertainty of $t_0$ corresponds to 90$\%$ probability of realization for the most affected case of the negation gate, motivating thus improvements in the experimental determination and control of the fast switching instant $t_0$. Finally, the condition $t>t_0$ strongly reduces the total probabilities of realization, since photons released before the time $t_0$ defined by the vertical dashed line in Fig.~\ref{fig:unary-gates} are lost. Rotations at late times therefore lead inevitably to significant losses, in the case of the logical negation, for instance, in average more than 80\% of the scattered photons. In the following we introduce two approaches that circumvent  the  depicted limitations.  

\begin{figure}%
\centering
\subfloat[time delay]{\label{fig:time}\includegraphics[width=0.42\linewidth]{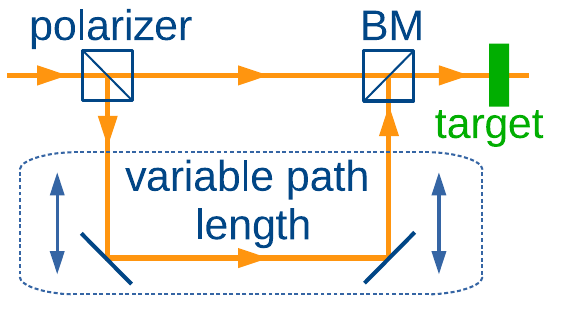}}%
\hspace{4pt}%
\subfloat[spatial splitting]{\label{fig:spatial}\includegraphics[width=0.42\linewidth]{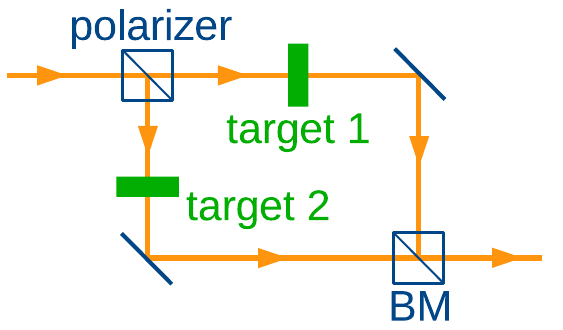}}%
\caption{(color online). The initial x-ray pulse can be temporally (a) or spatially (b) split  depending on its polarization.}
\label{fig:splitting}
\end{figure}

A first approach is to introduce a polarization-sensitive time delay line by using a polarizer as shown in Fig.~\subref*{fig:time}.  The instant of nuclear excitation becomes then dependent on the polarization of the incident SR pulse. Since the polarization is assumed to be either $\sigma$ or $\pi$ initially, the time delay can be chosen such that the two switching times exactly match and losses are minimized. For instance, in order to implement the logical negation via a switching time of 6.9 ns, the $\pi$-polarization needs to be delayed by 1.9 ns which reduces the losses from approx. 80\% to approx. 38\% in the case of initially $\pi$-polarized light. Practically, such a time delay line \cite{roling2012,osaka2014} can be realized with modern x-ray optics like channel-cut silicon crystals as polarizers \cite{Toellner1995.polarizer, Marx2011.polarizer, Marx2013.polarizer} and almost 100\% reflecting x-ray mirrors \cite{Shvydko2010.NP}. In Ref.~\cite{Roseker2011.time-delay, Roseker2013.time-delay}, for instance, 8 keV photons have been temporally delayed up to 3 ns. Similarly, in a second approach the polarizer can be used to spatially split the SR pulse in order to use two separated nuclear targets [see Fig.~\subref*{fig:spatial}]. The magnetic field rotations can be then chosen independently of each other. Analogously to the first approach, the switching  times can then be optimized individually for each input polarization state potentially leading to theoretical probabilities of realization larger than 97\%. The two spatially separated paths are later on recombined via a beam mixer BM \cite{Roseker2011.time-delay}.

\captionsetup[subfigure]{position=top}

We now turn to the implementation of binary logical gates by means of x-ray photons. Since the x-ray-nuclear interface hosts a single photon only,  a second, temporally synchronized photon is required in order to induce an effective nonlinearity as control. A simple but elegant idea is to have  the magnetic field rotation triggered by the second control photon. The latter can be spatially separated from the x-ray line as shown in Fig.~\ref{fig:setup} in order to distinguish between control and target mode. Since the nature of the implemented logical operation is completely determined by $t_0$, the trigger photon responsible for the magnetic field switching acts as control for the polarization-encoded x-ray. The magnetic field rotation could be applied at a predetermined switching time $t_0$ (counted from the incidence of the x-ray target pulse at $t=0$), in case  a control photon is detected at the trigger. Alternatively, the detection event of the control photon may trigger a prompt  rotation of the magnetic field. In this case the switching instant $t_0$ is no longer predetermined but rather set during operation, and coincides with  the incidence of the control photon at the trigger.

In the following, we show  how the canonical example of the controlled NOT (CNOT) gate  can be physically implemented with x-ray photons. A CNOT gate  flips the state of a target (T) qubit conditional on a control (C) qubit being in the logical state ``1'' \cite{NielsenChuang2000}. The control 
photon may encode information for instance in polarization, time bin \cite{Physics.time-bins, Humphreys2013.PRL, Donohue2013.PRL}, or path. Here, the NFS setup could be required to operate as identity or negation one-qubit gates for the target photon depending on the polarization state of the control photon. In order to render the triggering process  dependent on the polarization of the control photon, a polarization-sensitive element such as a polarizer can be used. 
The magnetic field rotation is applied with a predetermined switching time of 22.3~ns if a control photon is detected at the trigger.  With the same qubit notation as for the target x-ray photon, the polarizer  allows $\sigma$-polarized photons to reach the trigger, leading to the flip in polarization for the target photon. In this way, the photon detection at the trigger induces an effective interaction of control and target. Since the information associated with the control photon is destroyed during operation, and the polarization control relies on resonant scattering, the presented setup corresponds to a nondeterministic version of a destructive CNOT gate \cite{Pittman2001.PRA,Pittman2002.PRL}, which cannot be used directly for reversible computing \cite{OBrien2007.S}. However, provided copies of the target photon can be made, also a non-destructive CNOT gate can be accomplished. In the optical regime, this is achieved by harnessing quantum teleportation \cite{Bennett.1993.PRL}.

A proof-of-principle experiment can be carried out already today at SR facilities which have access to the keV photon energy regime and short pulses compared to the time-scale of the nuclear response ($\sim$ns). Moreover, the high pulse repetition rate renders it possible to record the presented intensity spectra in a reasonable time. The magnetic field rotation can be triggered by the detection of the control photon temporally synchronized with the synchrotron pulse clock. A fast triggering process is guaranteed by todays photodiodes which have response times shorter than 1~ns \cite{Seely2002.photodiode, Keister2010.photodiode}. A tilt of the polarization plane of the scattered x-rays can be measured with a precision down to a few arcsec by using a special polarizer-analyzer setup \cite{Marx2013.polarizer}. Experiments at novel x-ray free electron sources \cite{LCLS-web,SACLA-web}  will facilitate with their high brightness the implementation of binary x-ray gates where both target and control photons are resonant to the nuclear transitions and originate from the same x-ray pulse.

In conclusion, we have shown that the light-nucleus interaction in NFS can be used to perform logical operations on single x-ray photons  by applying  nuclear hyperfine magnetic field rotations. An x-ray photonic realization of single-qubit gates can be compiled by the mere variation of the magnetic field switching moment. An additional control qubit which triggers the magnetic field rotation can be exploited to design an x-ray  CNOT gate. The implementation of such basic logical operations with x-rays by using nuclear transitions may potentially advance quantum information in the near future towards new and promising parameter regimes characterized by long coherence times and sub-{\AA}ngstrom spatial resolution.

The authors would like to thank S. Tashenov for fruitful discussions.


\bibliographystyle{aipnum4-1}
\bibliography{mybibJ}{}

\end{document}